\documentclass{mem}
\usepackage{natbib}\usepackage{txfonts}\usepackage{balance}
\usepackage{graphicx}
\usepackage[a4paper]{hyperref}
\idline{75}{282}
\begin{document}
\def\teff{$T\rm_{eff }$}
\def\kms{$\mathrm {km s}^{-1}$}

\title{
Nature of X-shaped Radio Sources
}

   \subtitle{}

\author{
D.V. Lal,\inst{1,2}
M.J. Hardcastle,\inst{3} 
R.P. Kraft,\inst{4} 
C.C. Cheung,\inst{5} 
A.P. Lobanov,\inst{1} 
J.A. Zensus,\inst{1} 
S. Bhatnagar\inst{6} 
\and A.P. Rao\inst{2}
          }

  \offprints{D.V. Lal}

\institute{
Max-Planck-Institut f\"ur Radioastronomie,
Auf dem H\"ugel 69, 53121 Bonn, Germany
\mbox{\email{dharam@mpifr-bonn.mpg.de}}
\and
National Centre for Radio Astrophysics,
Pune University Campus, Pune 411007, India
\and
School of Physics, Astronomy, and Mathematics, University of Hertfordshire, Hatfield,~UK
\and
Harvard-Smithsonian Center for Astrophysics, MS-67, Cambridge, MA 02138, USA
\and
NASA GSFC,
Astrophysics Science Division, Greenbelt, MD 20771, USA
\and
National Radio Astronomy Observatory, Socorro, P.O. Box 0, New Mexico, USA
}

\authorrunning{Lal et~al. }

\titlerunning{Nature of X-shaped sources}

\abstract{
The nature of X-shaped sources is a matter of
considerable debate: it has even been proposed that they provide evidence for
black hole mergers$/$spin reorientation, and therefore constrain the rate of
strong gravitational wave events (Merritt \& Ekers 2002).
Based on morphological and spectral
characteristics of these sources, currently a strong contender to explain the
nature of these sources is the `alternative' model of Lal \& Rao (2007), in
which these sources consist of two pairs of jets, which are associated with two
unresolved AGNs.
Detailed morphological and spectral results on
milliarcsecond-scales (mas) provide a crucial test of this model, and hence
these sources are excellent candidates to study on mas; {\it i.e.}, to detect
the presence/absence of double nuclei/AGNs, signs of helical/disrupted jets,
thereby, to investigate spatially resolved/unresolved binary AGN systems and
providing clues to understanding the physics of merging of AGNs on mas.
We conducted a systematic study of a large sample of known X-shaped,
comparison FR\,II radio galaxies, and newly discovered X-shaped candidate
sources using Giant Metrewave Radio Telescope and Very Large Array
at several radio frequencies.
In our new observations of `comparison' FR\,II radio galaxies we find
that almost all of our targets show standard spectral steepening as a
function of distance from the hotspot.  However, one source, 3C\,321, has a
low-surface-brightness extension that shows a flatter
spectral index than the high-surface-brightness hotspots$/$lobes, as
found in `known' X-shaped sources.
}
\maketitle{}

\section{Introduction}

A small but significant fraction of low-luminosity radio galaxies
have a pair of large low-surface-brightness lobes oriented at an angle
to the high-surface-brightness `active' radio lobes, giving the total
source an `X' shape.
This peculiar and small subclass of extragalactic radio sources
is commonly referred to as X-shaped, or `winged' sources.
Typically, both sets of lobes are symmetrically aligned through
the centre of the associated elliptical host galaxy and are
approximately equal in linear extent.
Merritt \& Ekers (2002) noted that the majority of these sources
are FR\,IIs
and the rest are either FR\,Is or have an intermediate classification.

\subsection {Formation Scenario}

Several authors have attempted to explain the unusual
structure in X-shaped sources.
In some models these sources have been put forth as
derivatives of central engines that have been reoriented,
perhaps due to a minor merger (Merritt \& Ekers 2002;
Dennett-Thorpe et al. 2002; Gopal-Krishna et al. 2003).
Alternatively, they may also result from two pairs of jets
that are associated with a pair of unresolved AGNs (Lal \& Rao 2005, 2007).
These, however, are not the only interpretations for the unusual morphologies;
some authors suggest a hydrodynamic origin
(Worrall et al. 1995; Capetti et al. 2002;
Kraft et al. 2005) and some suggest a conical precession
of the jet axis (Rees 1978; Parma et al. 1985; Mack et al. 1994).
See Lal \& Rao (2007) and Cheung (2007) for a detailed account.

In several of the formation scenarios mentioned above for
X-shaped sources, the wings are interpreted as relics of past
radio jets and the active lobes as the newer ones.
Hence, the wings or low-surface-brightness features are expected to show
steeper spectra than the high-surface-brightness active lobes in
standard models for electron energy evolution (`spectral ageing').
Similarly, in typical FR\,II radio galaxies, the low-surface-brightness
features (regions away from the bright jet and hot-spot emission) are
expected to show steeper spectra than the bright jet and$/$or
the hot-spot emission. This can in principle be tested by observation.

Here, we present Giant Metrewave Radio Telescope (GMRT) and
Very Large Array (VLA) observations of
(i) the known sample of X-shapes sources, (ii) the control sample of
low-redshift {\it normal} FR\,II radio galaxies,
which is matched with the sample of known X-shaped radio sources
in size, morphological properties and redshift, and
(iii) a new sample selected from the candidate X-shaped sources (Cheung 2007).
Using observations carried out in the same
manner as those of Lal \& Rao (2007), we are able to investigate
whether standard spectral ageing models provide an adequate
description of sources in these samples, and discuss the
implications of our results for models of X-shaped sources.

In the paper, we define the spectral index $\alpha$ as
$S_\nu \propto \nu^\alpha$.

\begin{figure*}[t]
\begin{center}
\begin{tabular}{lll}
\includegraphics[width=4.1cm]{B1059_169_M240.PS} &
\includegraphics[width=4.1cm]{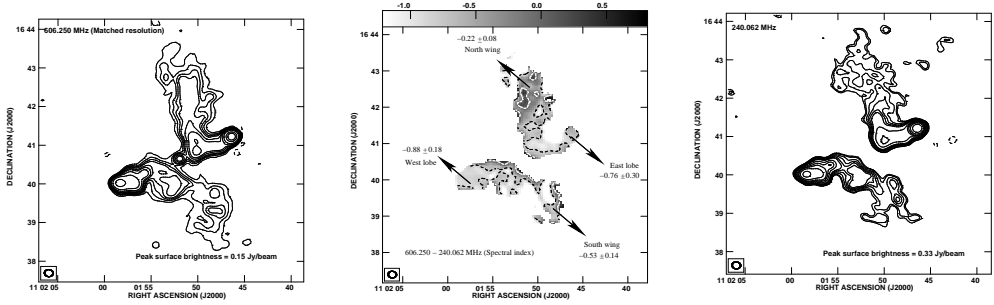} &
\includegraphics[width=4.1cm]{B1059_169_240.PS} \\ [-0.7cm]
\end{tabular}
\end{center}
\caption{
\footnotesize
GMRT map of B1059$+$169 (a known X-shaped source)
at 610 MHz (left panel), 240~MHz (right panel) and the spectral index
(240~MHz, 610~MHz) map (middle panel).
The 610~MHz map is matched with the resolution of 240~MHz.
The CLEAN beam for the matched maps is
13.0$^{\prime\prime}$~$\times$~11.1$^{\prime\prime}$
at a P.A. of 82.5$^{\circ}$.
}
\label{fig_x}
\end{figure*}

\section{Samples}

\noindent {\bf Known sample}
The earlier sample of known X-shaped sources was drawn from the list
compiled by Leahy \& Parma (1992).
There are nearly a dozen such sources, which
have been selected solely on the basis of their morphology,
and the sample is inhomogeneous and statistically incomplete.

\noindent {\bf Comparison sample}
The comparison sample consists of all nearby ($z < 0.1$)
normal FR~II sources from the 3CRR catalogue. These sources have radio
luminosities similar to those of the X-shaped sources, which lie close
to the FR~I/FR~II divide.  We impose an angular-size cutoff (based on
high-frequency radio maps) on the target sample and ensure that our
sample sources are of similar angular sizes to typical X-shaped sources.
In addition, all eight sample sources have known weak transverse extensions
(proto-wings?) and also have X-ray ({\it XMM}/{\it Chandra}) observations.

\noindent {\bf New sample}
The new sample is drawn from the compiled list
of nearly 100 new candidate X-shaped radio sources through a search of the
FIRST survey database (Cheung 2007).  Our sample sources have
(i)~characteristic `X' shape,
(ii) both set of lobes passing symmetrically through the
centre of the associated host galaxy, and
(iii) an angular size of more than 1.2$^{\prime}$ as seen in the VLA--FIRST
1.4~GHz images, which provided us with sixteen sources.

\section{GMRT Observations}

The GMRT has a hybrid configuration (Swarup et~al. 1991)
with 14 of its 30 antennas located in a central compact array
with size $\sim$1.1~km and the remaining antennas distributed in
a roughly `Y' shaped configuration, giving a maximum baseline length
of $\sim$25~km.
The 240 MHz and 610~MHz feeds of GMRT are
coaxial feeds enabling simultaneous multi-frequency observations
at these two frequencies.
We made synthesis observations of all sources in three samples at
240 MHz and 610 MHz, 
during several observing GTAC cycles, in the standard spectral line mode.
We also used archival data at 1.4~GHz from the VLA.

\begin{figure*}[t]
\begin{center}
\begin{tabular}{lll}
\includegraphics[width=4.6cm]{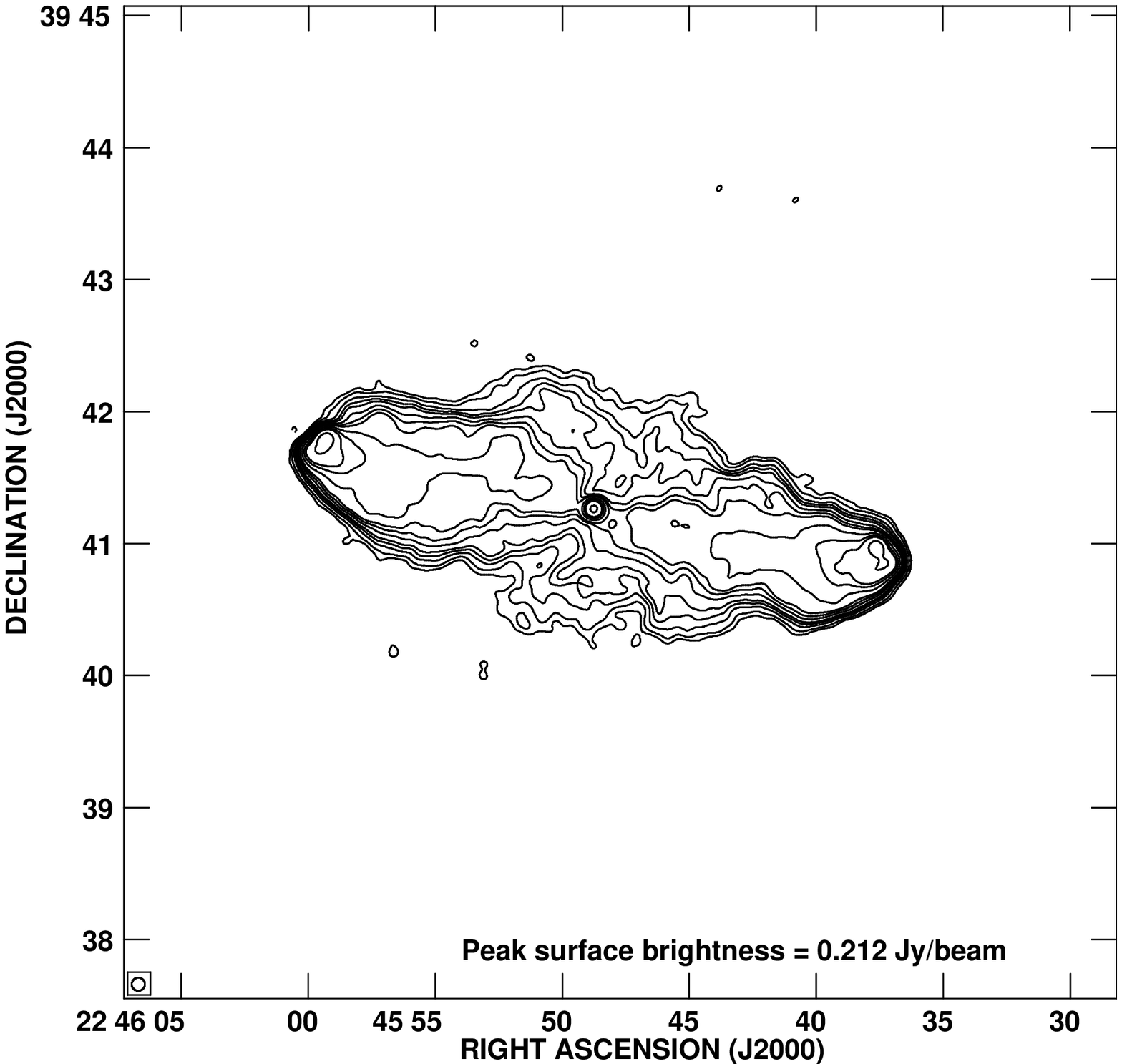} &
\hspace*{-0.2cm}\includegraphics[width=4.0cm]{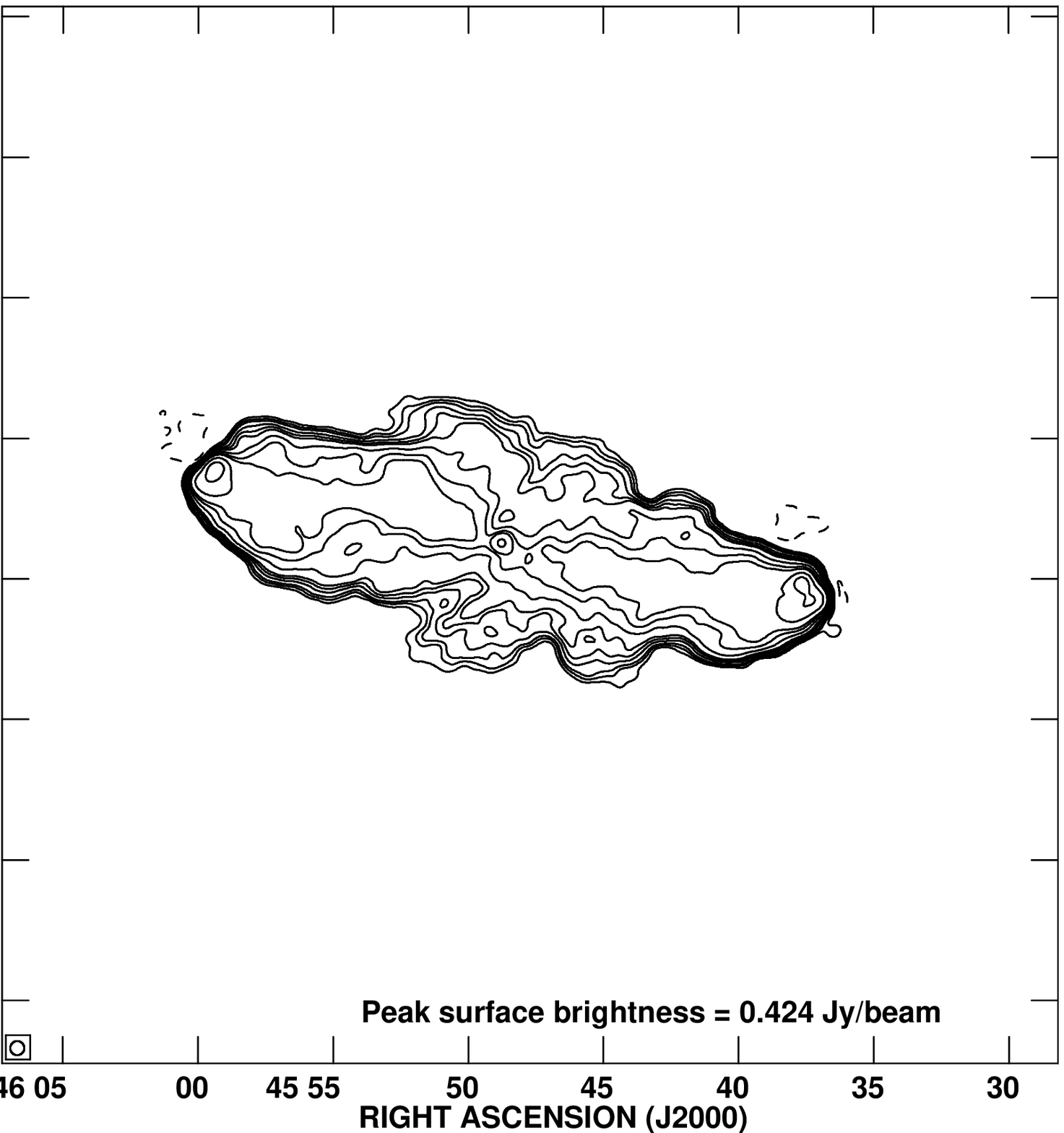} &
\hspace*{-0.2cm}\includegraphics[width=4.0cm]{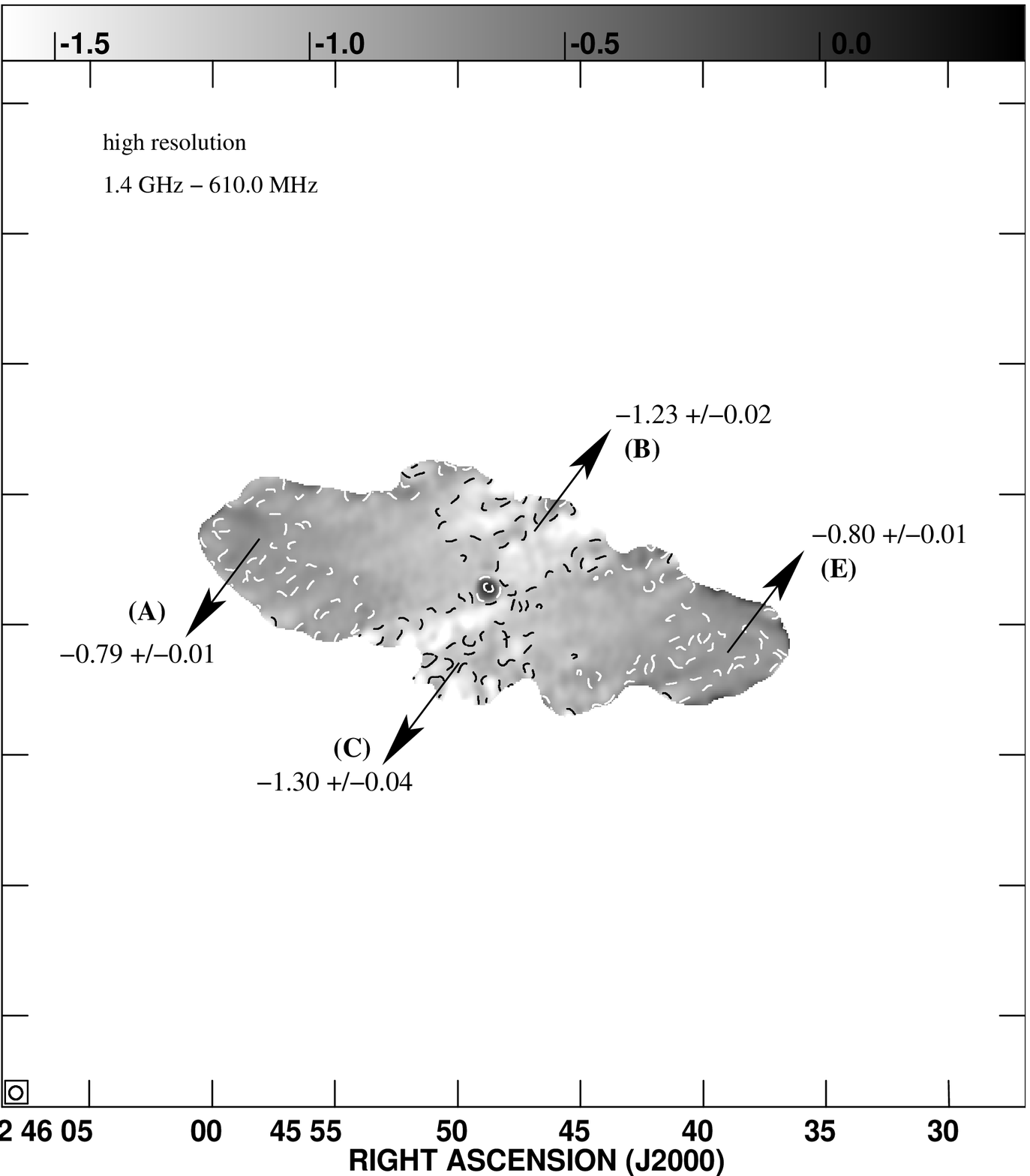} \\
\includegraphics[width=4.6cm]{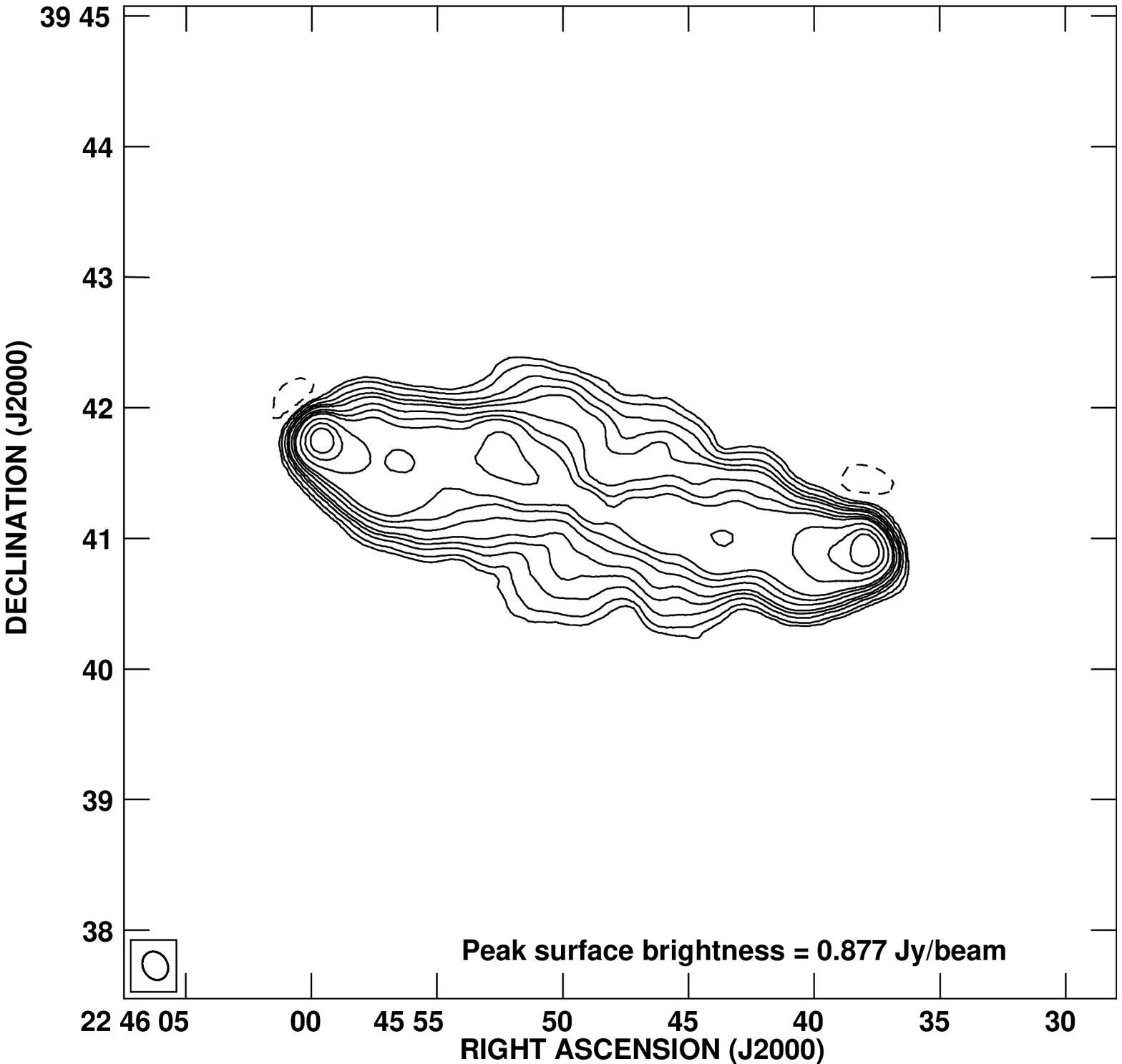} &
\hspace*{-0.2cm}\includegraphics[width=4.0cm]{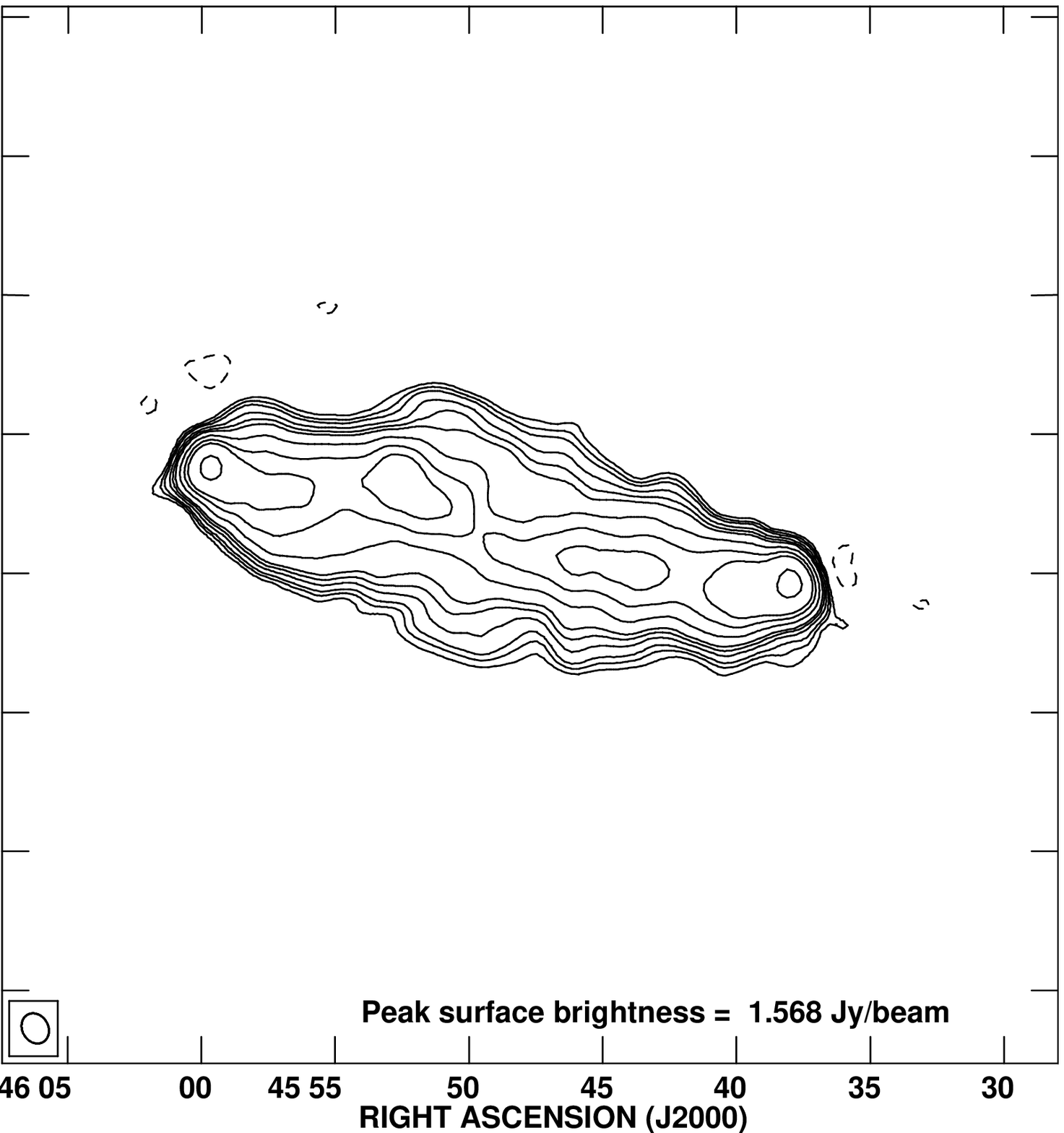} &
\hspace*{-0.2cm}\includegraphics[width=4.0cm]{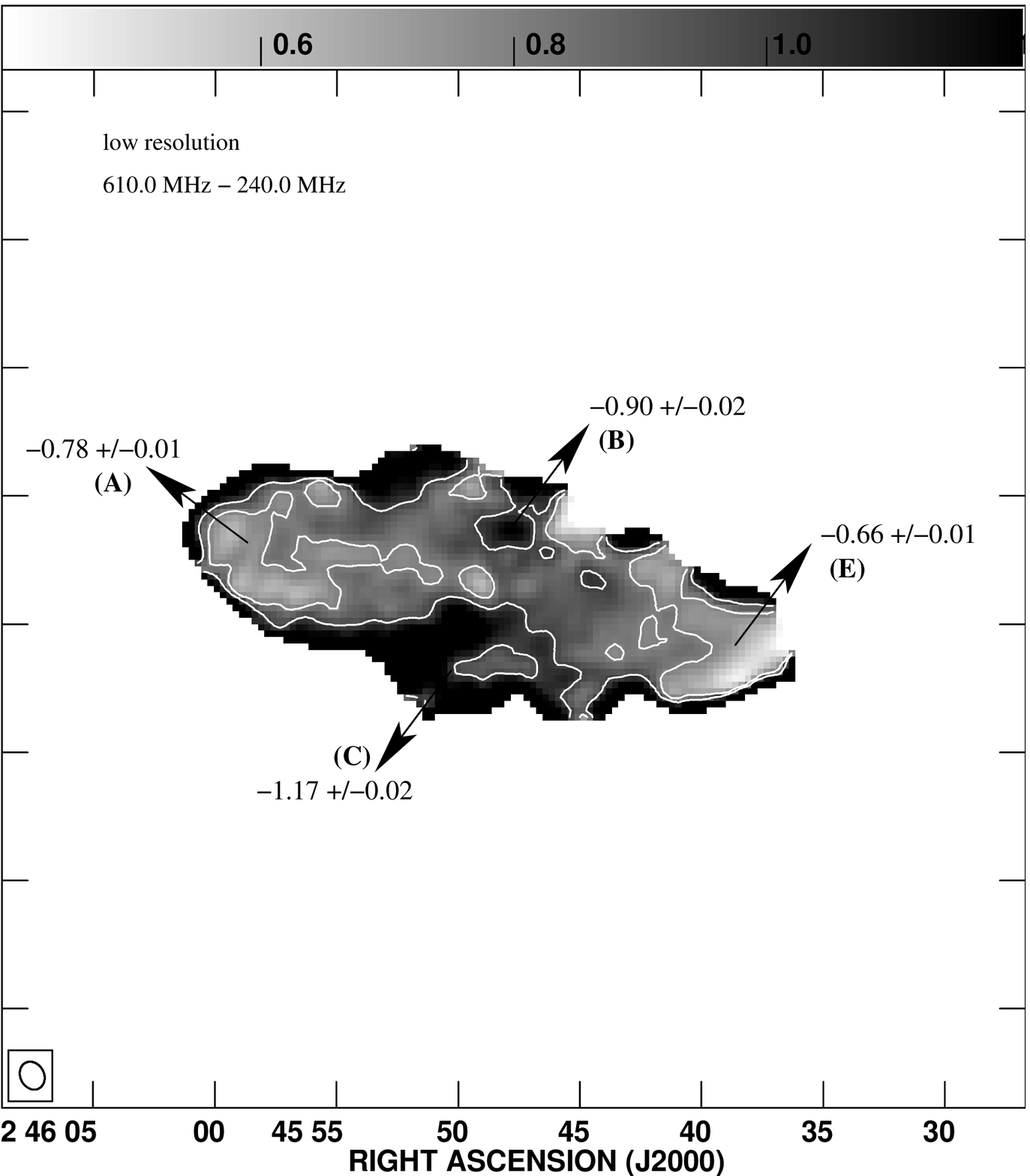} \\ [-0.4cm]
\end{tabular}
\end{center}
\caption{
\footnotesize
Images of 3C 452 (a FR~II sample source).
Upper left: The VLA map of 3C\,452 at 1.4 GHz.
Upper middle: The GMRT map of 3C\,452 at 610 MHz.
Lower left: The GMRT map of 3C\,452 at 610 MHz
matched with the resolution of 240~MHz.
Lower middle: The GMRT map of 3C\,452 at 240 MHz.
Upper right and lower right panels: The distribution of the spectral index,
between 1.4~GHz and 610 MHz (upper right),
and 240~MHz and 610 MHz (lower right), for the source.
The r.m.s. noise values in the radio images found at a source free location
are $\sim$0.2, $\sim$0.6 and $\sim$3.1~mJy~beam$^{-1}$ at
1.4~GHz, 610~MHz and 240~MHz, respectively.
The uniformly weighted CLEAN beams for upper and lower panel maps are
6.0$^{\prime\prime}$ $\times$~6.0$^{\prime\prime}$
at a P.A. of 0.0$^{\circ}$
and
13.6$^{\prime\prime}$ $\times$~11.2$^{\prime\prime}$
at a P.A. of $+$29.7$^{\circ}$, respectively.
}
\label{fig_frii}
\end{figure*}

\section{Results}

Fig.~\ref{fig_x} and~\ref{fig_frii} show examples of the images, which
have nearly complete ($u,v$) coverage, an angular resolution
$\sim$12$^{\prime\prime}$ and $\sim$6$^{\prime\prime}$ and the
r.m.s.~noise in the maps are $\sim$2.0 mJy~beam$^{-1}$ and
$\sim$0.2 mJy~beam$^{-1}$ at 240~MHz and 610~MHz, respectively.

The spectral characteristics of known X-shaped sources
seem to fall (nearly) equally into three distinct categories, namely, sources in which
(i) the wings have flatter spectral indices than the active lobes,
(ii) the wings and the active lobes have comparable spectral indices, and
(ii) the wings have steeper spectral indices than the active lobes.
It is probable that the three categories
of sources are unrelated to one another,
which makes it challenging to construct a single framework
explaining these spectral properties of X-shaped sources.
A strong contender~to~explain the
nature of these sources is the `alternative' model of Lal \& Rao (2007), in
which these sources consist of two pairs of jets associated with two
unresolved AGNs.

Almost all our `comparison sample' sources show monotonic steepening of the
radio spectrum from brighter hotspots to low surface brightness features, a
classical spectral signature seen in almost all normal FR\,II radio galaxies.
Preliminary analysis of the `new sample' also shows similar result.
In contrast to this,
a significant fraction of the `known sample' of X-shaped sources has wings with
flatter, or at least comparable, spectral indices to those in
the brighter active lobes.
This implies that the wings of X-shaped sources
do not simply behave like the low-surface-brightness
regions of more typical FR\,II sources.
The low surface brightness feature in 3C\,321, a classical FR\,II
radio source, also shows an unusual spectral behaviour, {\it i.e.},
the low-surface-brightness extension to one lobe shows a flatter
spectral index than the high-surface-brightness hotspots$/$lobes,
similar to the spectral
behaviour seen in wings in some of the X-shaped sources.
This raises the possibility that 3C\,321 consists of two
pairs of jets, which are associated with two unresolved AGNs, a
possible formation model for known X-shaped sources (Lal \& Rao 2007).
Another possibility is simply that our current understanding of
spectral ageing in radio lobes, \mbox{particularly at low frequencies, is incorrect.}

Further clues about the nature of X-shaped sources may be obtained from
milliarcsecond-scales (mas) study to search
for signs of binary AGN systems in X-shaped radio sources (Lobanov 2006)
either directly by spatially resolving the twin radio cores
or indirectly by studying the morphology of radio jets on
mas, and hence inferring their presence.

\begin{acknowledgements}
GMRT is run by the National Centre for Radio Astrophysics
of the Tata Institute of Fundamental Research.
The National Radio Astronomy Observatory is a
facility of the National Science Foundation operated under cooperative
agreement by Associated Universities, Inc.
\end{acknowledgements}

\bibliographystyle{aa}

\end{document}